\documentclass[sn-mathphys,Numbered]{sn-jnl}

\usepackage{graphicx}%
\usepackage{multirow}%
 \usepackage [latin1]{inputenc}
\usepackage{amsmath,amssymb,amsfonts}%
\usepackage{amsthm}%
\usepackage{mathrsfs}%
\usepackage[title]{appendix}%
\usepackage{xcolor}%
\usepackage{textcomp}%
\usepackage{manyfoot}%
\usepackage{booktabs}%
\usepackage{algorithm}%
\usepackage{algorithmicx}%
\usepackage{algpseudocode}%
\usepackage{listings}%

\theoremstyle{thmstyleone}%
\theoremstyle{thmstyletwo}%

\theoremstyle{thmstylethree}%

\begin{document}
	
	\title[A various equation of state for anisotropic models of compact star]{A various equation of state for anisotropic models of compact star}
	
	
	\author*[1]{\fnm{Rinkal} \sur{Patel}}\email{rinkalpatel22@gmail.com}
	
	\author[2]{\fnm{B. S.} \sur{Ratanpal}}\email{bharatratanpal@gmail.com}
	\equalcont{These authors contributed equally to this work.}
	
	\affil[1]{\orgdiv{Department of Applied Science \& Humanities}, \orgname{Parul University}, \orgaddress{\street{Limda}, \city{Vadodara}, \postcode{391 760}, \state{Gujarat}, \country{India}}}
	
	\affil[2]{\orgdiv{Department of Applied Mathematics}, \orgname{The Maharaja Sayajirao University of Baroda}, \orgaddress{\street{Faculty of Technology \& Engineering}, \city{Vadodara}, \postcode{390 001}, \state{Gujarat}, \country{India}}}
		
	
	\abstract{We obtain models of compact stars having pressure anisotropy on Finch-Skea spacetime by considering generalized equation of state (EoS), whose particular cases are linear, quadratic, polytropic, chaplygin and colour - flavor locked(CFL) equation of states. The physical viability of models are tested for strange star candidate 4U 1820 - 30 having mass $ M =  1.58M_\odot $ and radius R = 9.1 km. All the models are physically plausible.}

	\keywords{Equation of State, Einstein's field equation, Anisotropy}

	\maketitle
	\section{Introduction}\label{sec1}
	
	A compact star, as defined by general relativity, is a celestial object with a high density and a strong gravitational field. These stellar bodies, such as neutron stars and black holes, challenge our understanding of the cosmos and provide unique insights into the extreme nature of spacetime. General relativity is crucial in modelling and understanding the physics regulating compact stars, unravelling the secrets of their structure, behaviour and gravitational interactions. There are several models available in the literature describing relativistic compact objects.
	
	At extremely high densities, the stellar interior may experience asymmetrical radial and transverse stresses and the pressure inside the stellar object may be anisotropic in character \cite{ruderman1972pulsars}. Anisotropy may appear for a variety of reasons \cite{bowers1974anisotropic}. Numerous authors \cite{herrera1997local}, \cite{herrera1997thermal}, \cite{mak2002new}, \cite{mak2002exact}, \cite{chan2003gravitational}, \cite{harko2004anisotropy} have discussed how local anisotropy affects astrophysical objects and their causes. A solid core or the presence of type-3A superfluid \cite{harko2004anisotropy}, phase transitions during gravitational collapse \cite{sokolov1980phase}, \cite{herrera1989modeling}, pion condensation \cite{sokolov1980phase}, \cite{herrera1995jeans}, slow rotation of a fluid \cite{herrera1995jeans}, viscosity \cite{ivanov2010importance}, strong electromagnetic fields \cite{weber1999quark}, \cite{martinez2003magnetic}, \cite{usov2004electric}, etc. are some of the factors that cause anisotropy.
	
	Within the framework of general relativity, the EoS relates the energy density, pressure, and other thermodynamic parameters to the curvature of spacetime. This interaction between the EoS and general relativity is critical for adequately modelling compact objects such as neutron stars and black holes.
	\cite{feroze2011charged} investigated charged anisotropic matter with the quadratic EoS. By adopting a quadratic EoS relating radial pressure to energy density, \cite{maharaj2012regular} provide innovative correct solutions to the Einstein-Maxwell set of equations. \cite{sharma2013relativistic} investigated the analytic solution by assuming a certain radial pressure profile and utilising the Finch and Skea ansatz for the metric potential $g_{rr} $. \cite{malaver2014strange} used a quadratic EoS to describe the gravitational potential Z(x) of compact relativistic objects with anisotropic matter distributions. \cite{mafa2014stellar} model a charged anisotropic relativistic star with a quadratic EoS that includes the effect of the nonlinear term in the EoS. \cite{malaver2020relativistic} proposed new relativistic star configurations based on an anisotropic fluid distribution with a charge distribution. \cite{ivanov2001relativistic} investigated relativistic static fluid spheres with a linear EoS. \cite{maharaj2009generalized} examined the linear EoS for matter distributions with anisotropic pressures in the presence of an electromagnetic field using the metric potential $ g_{rr} = \frac{1}{a+bx^n}$. \cite{maharaj2014some} derived the solutions by considering charged anisotropic matter with a linear EoS $ p_r=\frac{1}{3} (\rho-\beta)$ consistent with quark stars. By selecting a particular form for one of the gravitational potentials and the electric field intensity,  \cite{ngubelanga2015compact} derived novel precise solutions in isotropic coordinates. Anisotropic compact stars on paraboloidal spacetime with linear EoS were investigated by \cite{thomas2017anisotropic}. \cite{ivanov2017analytical} studied analytical study of anisotropic compact star models that are similar to charged isotropic solutions. \cite{prasad2022anisotropic} used a linear EoS to study three different classes of innovative exact solutions for anisotropy factor. Recently \cite{patel2023new} investigated a new charged anisotropic solution on paraboloidal spacetime using a linear EoS that is compatible with a number of compact stars. 
	
	Polytropic EoS are useful in a wide range of astrophysical applications. \cite{binnington2009relativistic} used polytropic EoS to define electric-type and magnetic-type love numbers in the context of a spherical body affected by an external tidal field. \cite{chavanis2018simple} developed a simple universe model with a generalized EoS $ p = (\alpha + k\rho^{(1/n)} )\rho c^{2} $ that has a linear component $ p =\alpha \rho c^2 $ and a polytropic component $ p = k\rho^{(1+1/n)} c^2 $. \cite{takisa2013some} find accurate solutions for neutral anisotropic gravitating bodies in charged polytropic models. \cite{herrera2014conformally} investigated in depth conformally flat spherically symmetric fluid distributions that meet a polytropic EoS. \cite{ngubelanga2015relativistic} investigated the Einstein-Maxwell equation system in the framework of isotropic coordinates for anisotropic matter distributions in the presence of an electric field, assuming a polytropic EoS. \cite{herrera2016cracking} investigated the impact of modest fluctuations in local anisotropy of pressure and energy density on the incidence of cracking in spherical compact objects satisfying a polytropic EoS. \cite{azam2016study} studied the theory of newtonian and relativistic polytropes with generalized polytropic equations of state with anisotropic inner fluid distribution in the presence of charge. \cite{azam2017cracking} investigated the possibility of cracking in charged anisotropic polytropes with a generalized polytropic EoS under two alternative assumptions. Recently, \cite{nazar2023relativistic} proposed relativistic polytropic models of charged anisotropic compact objects.
	
	\cite{rahaman2010singularity} examined briefly singularity-free solutions for anisotropic charged fluids with the chaplygin EoS. \cite{bhar2018anisotropic} investigated closed-form solutions for modeling compact stars with interior matter distributions that obey a generalized chaplygin EoS. \cite{prasad2021behavior} proposed a new model of an anisotropic compact star in buchdahl spacetime that admits the chaplygin EoS. \cite{malaver2022analytical} investigated the analytical model of a compact star using a modified chaplygin EoS. 
		
	We have noted in \cite{nasheeha2021anisotropic} that authors have taken metric potential $ g_{rr}=\frac{1+ar^2}{1+(a-b)r^2}$ for various EoS quadratic, linear, polytropic, chaplygin, and color-flavor-locked EoS. It is noted that the metric potential $g_{tt}$ and many physical entities are not well-behaved in the case of  $ a = b $ for various EoS viz. quadratic, linear, polytropic, chaplygin, and color-flavor-locked EoS. We consider metric potential $ g_{rr} = 1+ar^2 $  which is particular case of $g_{rr} =\frac{1+ar^2}{1+(a-b)r^2}$  considered by \cite{nasheeha2021anisotropic} when $ a = b. $   We develop new models for anisotropic stars using a generalized version of nonlinear barotropic EoS with a specific gravitational potential $ g_{rr} $ and demonstrate how it can be reduced to other types of EoS to explain acceptable anisotropic matter distributions. Graphical analysis is done to investigate the physical acceptability of models.
	\section{The Field Equations}\label{sec2}
	The line element that characterizes the interior of a static, spherically symmetric compact star, is given by
	\begin{equation}\label{e1}
		ds^2 = e^{\nu(r)} dt^2-e^{\lambda(r)}dr^2 -r^2(d\theta^2 +sin^2 \theta d\phi^2).
	\end{equation}
	\\We take the energy-momentum tensor of the form
	\begin{equation}\label{e2}
		T_{ij}=(\rho+p_{\perp})u_iu_j+p_{\perp}g_{ij}+(p_r-p_{\perp})\chi_i\chi_j,
	\end{equation}
	where $ \rho $ is the matter density, $ p_{r} $ is the radial pressure, $ p_{\perp} $ is the tangential pressure, $u^{i}$ is the four-velocity of the fluid and $ \chi^{i}$ is a unit spacelike four-vector along the radial direction so that $u^{i}u_{i}= -1, \chi^{i}\chi_{j} = 1$ and $u^{i}\chi_{j}= 0$, with spacetime metric (\ref{e1}) and energy-momentum tensor (\ref{e2}), the Einstein's field equations takes the form		
	\begin{equation}\label{e3}
		8\pi\rho  = \frac{1-e^{-\lambda}}{r^2}+\frac{e^{-\lambda}\lambda'}{r},
	\end{equation} 
	\begin{equation}\label{e4}
		8\pi p_{r}  = \frac{e^{-\lambda}\nu'}{r}+\frac{e^{-\lambda}-1}{r^{2}},
	\end{equation}
	\begin{equation}\label{e5}
		8\pi p_{\perp} =e^{-\lambda} \left(\frac{\nu^{''}}{2} +\frac{\nu'^2}{4}-\frac{\nu' \lambda'}{4}+\frac{\nu'-\lambda'}{2r}\right),
	\end{equation}
	\begin{equation}\label{e6}
		8\pi \Delta = 8\pi P_{\perp}-8\pi P_{r}.
	\end{equation} 
	
	where primes denote differentiation with respect to r. The system of equation (\ref{e3}-\ref{e6}) governs the behavior of the gravitational field for an anisotropic fluid distribution.
	\section{Equation of State for Various Models}
	We consider a generalized equation of state of the form
	\begin{equation}\label{eos}
		p_r = \tau\rho^{(1+\frac{1}{p})}+\eta\rho-\omega,
	\end{equation}
	where $ \tau, \eta, \omega $ and p are real constants. If we put p=1 in equation (\ref{eos}), then it becomes quadratic EoS. If we put $\tau = 0 $ in equation (\ref{eos}), then it becomes linear EoS. If we fix $ \eta = 0, $ in equation (\ref{eos}), then it becomes polytrope with polytropic index p. If we set $ p = \frac{-1}{2}, \omega = 0 $ and $ \tau= -\alpha, $ in equation (\ref{eos}), then it becomes chaplygin EoS. If we set $p = -2, $ then it becomes color-flavor-locked (CFL)EoS.
	
	We solve the Einstein's field equations (\ref{e3}-\ref{e6}) together with EoS (\ref{eos}), to obtain an anisotropic model with EoS. For solving the system (\ref{e3}-\ref{e6}), we have three equations with five unknowns $(\rho, p_r, p_{\perp}, e^{\lambda}, e^{\nu}).$ We are free to select any two of them to complete this system. As a result, there are ten different ways to select any two unknowns. According to studies, \cite{bhar2015dark} choose $ \rho $ along with $ p_{r} $, \cite{murad2015some} and \cite{thirukkanesh2018anisotropic} select $ e^{\nu} $ with $ \Delta $ to model various compact stars \cite{sharma2013relativistic}, \cite{bhar2016anisotropic}, \cite{bhar2016new}  select $ e^{\lambda} $ and $ p_{r}. $ A frequent way, however, is to choose $ e^{\lambda} $  and EoS, which is a relationship between matter density and radial pressure $ p_r $. 
	\\To develop a physically reasonable model of the stellar
	configuration, we assume that the metric potential $ g_{rr} $ coefficient is expressed as  $ e^{\lambda} $ given by
	\begin{equation}\label{e7}
		e^{\lambda} = 1+a r^2,
	\end{equation}
	by selecting this metric potential, the function $ e^{\lambda} $ is guaranteed to be finite, continuous and well-defined within the range of stellar interiors.
	
	\begin{equation}
		\rho = \frac{a(3+ar^2)}{(1+ar^2)^2},
	\end{equation}
	\begin{equation}
		p_r = \tau\left(\left(\frac{a(3+ar^2)}{(1+ar^2)^2}\right)^{(1+\frac{1}{p})}-\left(\frac{a(3+aR^2)}{(1+aR^2)^2}\right)^{(1+\frac{1}{p})}\right) +\eta\left( \frac{a(3+ar^2)}{(1+ar^2)^2}- \frac{a(3+aR^2)}{(1+aR^2)^2}\right),
	\end{equation}
	\begin{equation}
		p_{\perp}=\frac{a}{4 \left(a r^2+1\right)^3}\left(f_1+f_2+\frac{ar^2(f_3+f_4)^2}{\left(a R^2+1\right)^4}+\frac{2(f_3+f_5+f_6)}{\left(a R^2+1\right)^2}\right),
	\end{equation}
	\begin{equation*}
		x=\left(\frac{a \left(a r^2+3\right)}{\left(a r^2+1\right)^2}\right)^{1/p},
	\end{equation*}
	\begin{equation*}
		y=\left(\frac{a \left(a R^2+3\right)}{\left(a R^2+1\right)^2}\right)^{1/p},
	\end{equation*}
	\begin{equation*}
		f_1= -\frac{4 a r^2 \tau  x \left(a r^2+5\right)}{p}+\frac{2 a \left(r^2 (-5 \eta +\tau  (x-6 y)+1)+R^2 (5 \eta +6 \tau  x-\tau  y+2)\right)2+6 \tau  (x-y)}{\left(a R^2+1\right)^2}-4(1+ a r^2),
	\end{equation*}
	\begin{equation*}
		f_2=\frac{2 a^2 \left(-3 r^4 (\eta +\tau  y)+2 r^2 R^2 (\tau  (x-y)+1)+R^4 (3 \eta +3 \tau  x+1)\right)-2 a^3 r^2 R^2 \left(r^2 (\eta +\tau  y)-R^2 (\eta +\tau  x+1)\right)}{\left(a R^2+1\right)^2},
	\end{equation*}
	\begin{equation*}
		f_3=a \left(r^2 (-5 \eta +\tau  x-6 \tau  y+1)+R^2 (5 \eta +6 \tau  x-\tau  y+2)\right)+3 \tau  (x-y)+1,
	\end{equation*}
	
	\begin{equation*}
		f_4=a^2 \left(-3 r^4 (\eta +\tau  y)+2 r^2 R^2 (\tau  x-\tau  y+1)+R^4 (3 \eta +3 \tau  x+1)\right)-a^3 r^2 R^2 \left(r^2 (\eta +\tau  y)-R^2 (\eta +\tau  x+1)\right),
	\end{equation*}
	\begin{equation*}
		f_5=a^4 r^4 R^2 \left(R^2 (\eta +\tau  x+1)-3 r^2 (\eta +\tau  y)\right)+a^3 \left(-9 r^6 (\eta +\tau  y)+r^4 R^2 (-5 \eta +2 \tau  x-7 \tau  y+2)+2 r^2 R^4\right),
	\end{equation*}
	\begin{equation*}
		f_6=a^2 \left(r^4 (-20 \eta +\tau  x-21 \tau  y+1)-r^2 R^2 (5 \eta +5 \tau  y-4)+R^4 (3 \eta +3 \tau  x+1)\right).
	\end{equation*}
	
	\begin{equation}\label{e8}
		e^{\nu} = C \;(1+ar^2)^\eta exp{(f_7+f_8)},
	\end{equation}
	\begin{equation*}
		f_7=\frac{p \tau  x \left(a r^2+1\right)  Hypergeometric \; _2F_1\left(\frac{1}{p}-1,-\frac{1}{p};\frac{1}{p};-\frac{2}{a r^2+1}\right)}{2(p-1)\left(\frac{2}{a r^2+1}+1\right)^{1/p}}-\frac{\left(a r^2+1\right)^2 \left(a R^2+3\right) (\eta +\tau  y)}{4\left(a R^2+1\right)^2},
	\end{equation*}
	\begin{equation*}
		f_8=\frac{(\eta +1) \left(a r^2+1\right)}{4}-\frac{ \text{p} \tau  Hypergeometric \; _2F_1\left(-\frac{1}{p},\frac{1}{p};1+\frac{1}{p};-\frac{2}{a r^2+1}\right)}{\left(\frac{2}{a r^2+1}+1\right)^{1/p} }.
	\end{equation*}
	
	where, C is a constant of integration.
	\begin{equation}
		C =\frac{1}{(1+aR^2)^{(\eta+1)}}exp\left(\frac{\left(a R^2+3\right) (\eta +\tau  y)-2 (\eta +1) \left(a R^2+1\right)}{4}+\frac{2(p-1)p \tau  yf_9}{2 (p-1)\left(\frac{2}{a R^2+1}+1\right)^{1/p}}\right),
	\end{equation}
	\begin{equation*}
		f_9=Hypergeometric \; _2F_1\left(-\frac{1}{p},\frac{1}{p};1+\frac{1}{p};-\frac{2}{a R^2+1}\right)
	\end{equation*}
	\begin{equation*}
		-\left(a R^2+1\right) Hypergeometric \; _2F_1\left(\frac{1}{p}-1,-\frac{1}{p};\frac{1}{p};-\frac{2}{a R^2+1}\right).
	\end{equation*}
	
	The mass function within the sphere of radius R for the
	metric potential equation (\ref{e7}) is given by
	\begin{equation}
		M=m(r) = \frac{aR^3}{2(1+aR^2)}.
	\end{equation}
	To study the physical behavior of realistic stars, the closed-form solutions. In our generated model, equation (\ref{e8}) is the solution for different values of p. Subsequently, we consider the cases $ p=2 $  and $ p=-\frac{1}{2} $  in the
	following sections which are of physical interest.
	\subsection{ Quadratic Equation of State}
	If we set p = 1 then equation (\ref{eos}) takes the form of quadratic EoS
	\begin{equation}\label{e9}
		p_r = \tau\rho^{2}+\eta\rho-\omega,
	\end{equation}
	\begin{equation}
		e^{\lambda} = 1+a r^2,
	\end{equation}
	\begin{equation}
		\rho = \frac{a(3+ar^2)}{(1+ar^2)^2},
	\end{equation}
	\begin{equation}
		p_r = \tau\left(\left(\frac{a(3+ar^2)}{(1+ar^2)^2}\right)^{2}-\left(\frac{a(3+aR^2)}{(1+aR^2)^2}\right)^{2}\right) +\eta\left( \frac{a(3+ar^2)}{(1+ar^2)^2}- \frac{a(3+aR^2)}{(1+aR^2)^2}\right),
	\end{equation}
	\begin{equation}
		p_{\perp}= \frac{a}{4 \left(a r^2+1\right)}\left( f_{10}+f_{11}+f_{12}+f_{13}\right), 
	\end{equation}
	\begin{equation*}
		f_{10}=	\frac{16 a \tau-2 a r^2 (a \tau +2 \eta ) }{\left(a r^2+1\right)^2}-\frac{2 a r^2+4+2 a \eta  r^2-(4 a \tau +8 \eta)}{a r^2+1}+4 (\eta+1),
	\end{equation*}
	\begin{equation*}
		f_{11}=(2 a r^2-4 \left(a r^2+1\right)-8 a^2 \tau  r^2)p_{14}+\frac{16 a \tau-8 a^2 \tau  r^2 }{\left(a r^2+1\right)^3},
	\end{equation*}
	
	\begin{equation*}
		f_{12}	=4 a r^2 \left(-f_{14}-\frac{a \tau +2 \eta }{\left(a r^2+1\right)^2}-\frac{8 a \tau }{\left(a r^2+1\right)^3}-\frac{12 a \tau }{\left(a r^2+1\right)^4}\right),
	\end{equation*}
	\begin{equation*}
		f_{13}	=	a r^2 \left((-a r^2+1)f_{14}+\frac{a \tau +2 \eta }{a r^2+1}+\frac{4 a \tau }{\left(a r^2+1\right)^2}+\frac{4 a \tau }{\left(a r^2+1\right)^3}+\eta +1\right)^2,
	\end{equation*}
	\begin{equation*}
		f_{14}=\frac{\left(a R^2+3\right) \left(a^2 R^2 \left(\tau +\eta  R^2\right)+a \left(3 \tau +2 \eta  R^2\right)+\eta \right)}{\left(a R^2+1\right)^4},
	\end{equation*}
	\begin{equation}\label{enuQ}
		e^{\nu}=C\;(1+ar^2)^{(\frac{a\tau+2\eta}{2})}exp{\left( f_{15}\right) },
	\end{equation}
	\begin{equation*}
		f_{15}= -\frac{a\tau}{(1+ar^2)^2}-\frac{2a\tau}{(1+ar^2)}+\frac{(1+ar^2)(1+\eta)}{2}
	\end{equation*}
	\begin{equation*}
		-\frac{(3+aR^2)(1+ar^2)^2\left(\eta+a^2R^2(\tau+R^2\eta)+a(3\tau+2R^2\eta)\right)}{4(1+aR)^4},
	\end{equation*}
	
	where,
	\begin{equation}
		C=\frac{exp\left( -\frac{a\tau}{(1+aR^2)^2}-\frac{2a\tau}{(1+aR^2)}+\frac{(1+aR^2)(1+\eta)}{2}-\frac{(3+aR^2)\left(\eta+a^2R^2(\tau+R^2\eta)+a(3\tau+2R^2\eta)\right)}{4(1+aR)^2}\right) }{(1+aR^2)^{(\frac{a\tau+2\eta}{2}+1)}}.
	\end{equation}
	The same metric potential is used by \cite{sharma2013relativistic}, but assuming radial pressure in the form of $8\pi p_r=\frac{p_0(1-\frac{r^2}{R^2})}{R^2(1+\frac{r^2}{R^2})^2}.$ Similarly \cite{feroze2011charged} used same ansatz for metric potential $g_{rr}$ considering  charged anisotropic matter with quadratic
	equation of state.
	\subsection{Linear Equation of State}
	If we set $ \tau = 0$, equation(\ref{eos}) becomes
	\begin{equation}\label{e9*}
		p_r = \eta\rho-\omega,
	\end{equation}
	\begin{equation}
		e^{\lambda} = 1+ar^2,
	\end{equation}
	\begin{equation}
		\rho = \frac{a(3+ar^2)}{(1+ar^2)^2},
	\end{equation}
	\begin{equation}
		p_r = \eta\left( \frac{a(3+ar^2)}{(1+ar^2)^2}- \frac{a(3+aR^2)}{(1+aR^2)^2}\right),
	\end{equation}
	\begin{equation*}
		p_{\perp}= \frac{a\left(4(1+\eta)+\frac{8\eta-4-2ar^2(1+\eta)}{1+ar^2}+2 \eta  \left(-\frac{a r^2 \left(a R^2+3\right)}{\left(a R^2+1\right)^2}-\frac{2 \left(a r^2+1\right) \left(a R^2+3\right)}{\left(a R^2+1\right)^2}-\frac{6 a r^2}{\left(a r^2+1\right)^2}\right)\right)}{4(1+ar^2)}
	\end{equation*}
	\begin{equation}
		+	\frac{a r^2 \left(\frac{2 \eta }{a r^2+1}-\frac{\eta  \left(a r^2+1\right) \left(a R^2+3\right)}{\left(a R^2+1\right)^2}+\eta +1\right)^2}{4(1+ar^2)},
	\end{equation}
	\begin{equation}\label{enuL}
		e^{\nu}=C\;(1+ar^2)^{\eta}exp{\left(  \frac{(1+ar^2)(1+\eta)}{2}-\frac{\eta (3+aR^2)(1+ar^2)^2}{(1+aR^2)^2}\right)  },
	\end{equation}
	where,
	\begin{equation}
		C=\frac{exp\left( \frac{-2+\eta-aR^2(2+\eta)}{4}\right) }{(1+aR^2)^{(\eta+1)}}.
	\end{equation}
	which is similar to the solution of \cite{thomas2017anisotropic} with $ a= \frac{1}{L^2} $.
	\subsection{Polytropic Equation of State}
	When p = 2 and $\eta = 0$, equation (\ref{eos}) takes a polytropic form
	\begin{equation}
		p_r=\tau\rho^{(3/2)}-\omega,
	\end{equation}
	so Einstein's field equations become,
	\begin{equation}
		p_r = \tau\left( \left(\frac{a(3+ar^2)}{(1+ar^2)^2}\right)^{(3/2)}- \left(\frac{a(3+aR^2)}{(1+aR^2)^2}\right)^{(3/2)}\right),
	\end{equation}
	\begin{equation}
		p_{\perp}= \frac{a\left(-2 \left(a r^2+1\right)^3 \left(a R^2+1\right)^2f_{16}- a r^2 \left(a r^2+1\right)^2 f_{17}\left(2\left(a R^2+1\right)^2-f_{17}\right)+\frac{2 \left(a R^2+1\right)^2}{l}f_{18}\right)}{4 \left(a r^2+1\right)^5 \left(a R^2+1\right)^4},
	\end{equation}
	
	\begin{equation*}
		l= \sqrt{\frac{a \left(a r^2+3\right)}{\left(a r^2+1\right)^2}}, \;\; m= \sqrt{\frac{a \left(a R^2+3\right)}{\left(a R^2+1\right)^2}},
	\end{equation*}
	\begin{equation*}
		f_{16}=a^3 \left(m r^4 R^2 \tau -r^2 R^4 (l \tau +1)\right)+a^2 \left(-2 r^2 R^2 (l \tau -m \tau +1)+R^4 (1-3 l \tau )+3 m r^4 \tau \right)
	\end{equation*}
	\begin{equation*}
		-3 l \tau +3 m \tau +1-a \left(r^2 (l \tau -6 m \tau +1)+R^2 (6 l \tau -m \tau -2)\right),
	\end{equation*}
	\begin{equation*}
		f_{17}=a^3 \left(r^2 R^4 (l \tau +1)-m r^4 R^2 \tau \right)+a^2 \left(2 r^2 R^2 (l \tau -m \tau +1)+R^4 (3 l \tau +1)-3 m r^4 \tau \right)+3 \tau  (l-m)+1
	\end{equation*}
	\begin{equation*}
		+a \left(r^2 (l \tau -6 m \tau +1)+R^2 (6 l \tau -m \tau +2)\right),
	\end{equation*}
	\begin{equation*}
		f_{18}=a^6 l r^8 R^2 \left(R^2-3 m r^2 \tau \right)+a^5 \left(-9 l m r^{10} \tau +l r^8 R^2 (2-13 m \tau )+4 l r^6 R^4-5 r^4 R^4 \tau \right)+l (1-3 m \tau )
	\end{equation*}
	\begin{equation*}
		+a^4 \left(l r^8 (1-39 m \tau )+2 l r^6 R^2 (4-11 m \tau )+2 r^4 \left(3 l R^4-5 R^2 \tau \right)-12 r^2 R^4 \tau \right)
	\end{equation*}
	\begin{equation*}
		+a^2 \left(6 l r^4 (1-9 m \tau )+r^2 \left(l R^2 (8-7 m \tau )-12 \tau \right)+l R^4+18 R^2 \tau \right)+a \left(l r^2 (4-21 m \tau )+l R^2 (2-m \tau )+9 \tau \right)
	\end{equation*}
	\begin{equation*}
		+a^3 \left(2 l r^6 (2-33 m \tau )+r^4 \left(-6 l R^2 (3 m \tau -2)-5 \tau \right)+4 r^2 \left(l R^4-6 R^2 \tau \right)+9 R^4 \tau \right),
	\end{equation*}
	
	\begin{equation}\label{enuP}
		e^{\nu}=C\; exp{\left(  \frac{f_{19}}{4 \left(a R^2+1\right)^2}\right)  },
	\end{equation}
	\begin{equation*}
		f_{19}= 2-3m\tau+a^3 r^2 R^2(-r^2m\tau+R^2(2+4l\tau))+a(r^2(2+4l\tau-6m\tau)+R^2(4-m\tau))
	\end{equation*}
	\begin{equation*}
		+a^2(2R^4-3r^4m\tau+2r^2R^2(2+4l\tau-m\tau)-6 \sqrt{2} \sqrt{a} \tau  \left(a R^2+1\right)^2 \tanh ^{-1}\left(\frac{\left(a r^2+1\right) m}{\sqrt{2} \sqrt{a}}\right),
	\end{equation*}
	where,
	\begin{equation*}
		C = \frac{\exp \left(-\frac{a^3 R^6 \left(3 \tau  m+2\right)+a^2 R^4 \left(3 \tau  m+6\right)+\text{aR}^2 \left(6-3 \tau  m\right)-3 \tau m -6 \sqrt{2} \sqrt{a} \tau  \left(a R^2+1\right)^2 \tanh ^{-1}\left(\frac{\left(a R^2+1\right) m}{\sqrt{2} \sqrt{a}}\right)+2}{4 \left(a R^2+1\right)^2}\right)}{a R^2+1}.
	\end{equation*}
	This is the new solution for this gravitational potential in polytropic EoS.
	\subsection{Chaplygin Equation of State}
	If we set $ p= -\frac{1}{2}, \tau = -\alpha $  and $ \omega =0 $ then equation (\ref{eos}) becomes Chaplygin EoS as
	\begin{equation}
		p_r = \eta\rho - \frac{\alpha}{\rho},
	\end{equation}
	\begin{equation}
		p_{\perp}=\frac{f_{20}+f_{21}+f_{22}}{4 a^2 \left(a r^2+1\right)^3 \left(a r^2+3\right)^2},
	\end{equation}
	
	\begin{equation*}
		f_{20}= a^8 r^{10} \left(\eta -\alpha  r^4+1\right)^2+2 a^7 r^8 \left(6 \eta ^2+11 \eta +4 \alpha ^2 r^8-\alpha  (10 \eta +13) r^4+5\right)
	\end{equation*}
	\begin{equation*}
		+2 a^2 \alpha  r^2 \left(-9 \eta +14 \alpha  r^4-48\right)+4 a \alpha  \left(2 \alpha  r^4-3\right)+\alpha ^2 r^2,
	\end{equation*}
	\begin{equation*}
		f_{21}= 2 a^6 r^6 \left(27 \eta ^2+43 \eta +14 \alpha ^2 r^8-3 \alpha  (13 \eta +22) r^4+18\right)+2 a^3 \left(54 \eta +28 \alpha ^2 r^8-3 \alpha  (14 \eta +47) r^4\right),
	\end{equation*}
	\begin{equation*}
		f_{22}= 2 a^5 r^4 \left(54 \eta ^2+75 \eta +28 \alpha ^2 r^8-4 \alpha  (19 \eta +40) r^4+27\right)
	\end{equation*}
	\begin{equation*}
		+a^4 r^2 \left(9 \left(9 \eta ^2+16 \eta +3\right)+70 \alpha ^2 r^8-2 \alpha  (79 \eta +205) r^4\right),
	\end{equation*}
	\begin{equation}\label{enuC}
		e^{\nu}= C\; (1+ar^2)^{\eta}(3+ar^2)^{\frac{4\alpha}{a^2}}exp\left( \frac{-(1+ar^2)(10\alpha-ar^2\alpha+a^2(-3+r^4\alpha-3\eta))}{6a^2}\right) ,
	\end{equation}
	where,
	\begin{equation}
		C = \frac{(3+aR^2)^{(\frac{-4\alpha}{a^2})}exp\left( -(\frac{-(1+aR^2)(10\alpha-aR^2\alpha+a^2(-3+R^4\alpha-3\eta))}{6a^2})\right) }{(1+aR^2)^{(\eta+1)}}.
	\end{equation}
	
	\subsection{CFL Equation of State}
	When the internal structure of a compact star formed of strange matter is in the CFL phase, CFL EoS is used.
	If we set $ p = -2 $, then equation (\ref{eos}) becomes CFL EoS (\cite{thirukkanesh2020comparative}, \cite{rocha2019exact})
	\begin{equation}
		p_r = \tau\rho^{\frac{1}{2}}+\eta\rho-\omega,
	\end{equation}
	\begin{equation}
		p_{\perp} =-\frac{f_{23}+f_{24}+f_{25}+f_{26}}{36 \left(a r^2+1\right)},
	\end{equation}
	\begin{equation*}
		f_{23}=\frac{108 a^2 \eta  r^2}{\left(a r^2+1\right)^2}+\frac{18a\eta \left(a R^2+3\right)\left(ar^2\left(a r^2+1\right)\right)}{\left(a R^2+1\right)^2}-36a (\eta +1+ r^2 \tau p) -48 \tau  \left(a r^2+2\right)p,
	\end{equation*}
	
	\begin{equation*}
		f_{24}=\frac{18 a^2 (\eta +1) r^2+24 a r^2 \tau  \left(a r^2+2\right)+36 a(1-2\eta)-6 a r^2 \tau p \left(a r^2+5\right)}{a r^2+1},
	\end{equation*}
	\begin{equation*}
		f_{25}=\frac{6a^2 r^2 \tau \left(a r^2+5\right)\left(4\left(a r^2+2\right) - \left(a r^2+5\right)\right)}{\left(a r^2+1\right)^3 p}+18q\tau(3ar^2+2)+12 \tau  \left(a r^2+5\right) p,
	\end{equation*}
	
	\begin{equation*}
		f_{26}=r^2 \left(-3 a (\eta +1)-\frac{6 a \eta }{a r^2+1}+3q(1+ar^2)(q\eta+\tau)-4 \tau  \left(a r^2+2\right) p+\tau  \left(a r^2+5\right)p\right)^2,
	\end{equation*}
	
	\begin{equation*}
		e^{\nu}=C\;(1+ar^2)^{\eta}
	\end{equation*}
	\begin{equation}\label{enucf}
		exp\left(  \frac{ \tau  \left(a^2 r^4+4 a r^2+3\right)p}{3 a}+\frac{2 (\eta +1) \left(a r^2+1\right)}{4}-\frac{\eta  \left(a r^2+1\right)^2 \left(a R^2+3\right)}{4\left(a R^2+1\right)^2}-\frac{\tau  \left(a r^2+1\right)^2 q}{4a}\right)  ,
	\end{equation}
	where,
	\begin{equation}
		C = \frac{exp\left(  -\frac{ \tau q \left(a^2 R^4+4 a R^2+3\right)}{3 a}-\frac{(\eta +1) \left(a R^2+1\right)}{2}+\frac{\eta  \left(a R^2+3\right)}{4}+\frac{p\tau\left(a R^2+1\right)^2}{4a}\right)  }{(a R^2+1)^{(\eta+1)}}.
	\end{equation}
	All of the physical plausibility conditions are representing in the next section.

	\section{Conditions for physical acceptability}\label{sec4}
	Which of the following for the model to be physically realistic star(\cite{mak2002exact}, \cite{thirukkanesh2012exact}, \cite{delgaty1998physical}).
	\\	a) Regularity condition. 
	\\i) Gravitational potentials should be regular at the center of the star.
	\\ii) $ \rho(r) \ge 0 ,\;\;\;\;p_{r}(r)\ge 0 ,\;\;\;\;p_{\perp}(r)\ge 0 $ \;\;\;\; for  $ 0 \le r \le R.$
	\\	iii) $ p_{r}(r=R) = 0 $
	
	b) Junction condition: the interior metric (1) must be matched smoothly at the boundary
	(r = R) with Schwarzschild exterior metric
	\begin{equation}\label{junction}
		ds^{2}=\left(1-\frac{2M}{R}\right)dt^{2}-\left(1-\frac{2M}{R}\right)^{-1}dr^{2} -r^2(d\theta^2+sin^2\theta d\phi^2),
	\end{equation} 
	at the boundry $ r=R $ of star.
	
	c) Behavior of measure of anisotropy.
	
	d) Causality condition:
	\;\;\;\; $0 \le \frac{dp_{r}}{d\rho} \le 1$ and $0 \le \frac{dp_{\perp}}{d\rho} \le 1 $ \;\;\;\;\;\; for  $ 0 \le r \le R  $
	
	e) Not cracking condition: for the stability of a star it is required to satisfy the
	condition  $ -1 \le \frac{dp_{\perp}}{d\rho}-\frac{dp_{r}}{d\rho} \le 1.$
	
	f) Energy condition:
	\;\;\;\;$ \rho - p_{r} - 2 p_{\perp}\ge 0.$ \;\;\;\;\;\; for  $ 0 \le r \le R  $
	
	g) Monotony condition:
	\;\; $\frac{d\rho}{dr}\le 0 ,\;\;  \frac{dp_{r}}{dr} \le 0 ,\;\;   \frac{dp_{\perp}}{dr}\le 0  $ \;\;\; for  $ 0 \le r \le R  $
	
	h) Gravitational redshift and surface redshift. 
	
	i) Stability under three different forces.
	
	j) Adiabatic index for stability.
	
	\begin{table}[h]
		\caption{The numerical values of the $\tau, \eta, \omega $ and constant(a)  for the compact star 4U 1820-30.}
		\label{tab:1}
		\begin{tabular}{lllllll}
			\hline\noalign{\smallskip}
			\textbf{Equation of states} & $ \mathbf{ \tau } $ & $ \mathbf{ \eta } $ & $ \mathbf{ \omega } $ &  $ \mathbf{ a } $ & $ \mathbf{\rho(0)} $ & $ \mathbf{\rho(R)} $ \\
			& & & & & \textbf{(MeV fm{$\mathbf{^{-3}}$})} & \textbf{(MeV fm{$\mathbf{^{-3}}$})}
			\\\textbf{$ Quadratic $} 	  & 0.1  & 0.15   &    1   & 0.01 & 903.407 & 344.942\\
			\textbf{$ Linear $} 	   & 0  & 0.15   &    1   & 0.01 & 903.407 & 344.942 \\
			\textbf{$ Polytropic $} 	   & 1.5  & 0   &    0   & 0.01 & 903.407 & 344.942 \\
			\textbf{$ Chaplygin $}     & 0.1  & 0.1   &    0   & 0.01 & 903.407 & 344.942 \\
			\textbf{$ CFL $}    & 0.01  & 0.15   &    0   & 0.01 & 903.407 & 344.942 \\
			
			\noalign{\smallskip}\hline
			
		\end{tabular} 
		
	\end{table}
	\begin{table}[h]
		\caption{The numerical values of the strong energy condition at the center as well as surface, redshift at the surface and adiabatic Index at the surface for the compact star 4U 1820-30.}
		\label{tab:2}       
		\begin{tabular}{llllll} 
			\hline\noalign{\smallskip}
			\textbf{Equation of states} & 
			{$ \mathbf{ \rho - p_{r} - 2p_{\perp}}_{(r=0)} $} & {$ \mathbf{\rho-p_{r}-2p_{\perp}}_{(r=R)} $} &    {$ \mathbf{ Z_{(r=0)}} $} &    {$ \mathbf{ Z_{(r=R)}} $} &  {$ \mathbf{\Gamma_{(r=0)}}$} 
			
			\\	&   \textbf{} & \textbf{} &    \textbf{(Redshift)} &    \textbf{(Redshift)} & \textbf{(Adiabatic }   \\
			& \textbf{}	&  \textbf{} & \textbf{} & \textbf{} &\textbf{ Index)}   \\
			\noalign{\smallskip}\hline\noalign{\smallskip}
			\textbf{$ Quadratic $} 	  & 645.152  & 274.575 &    0.707848   & 0.352072 &1.79312 \\
			\textbf{$ Linear $} 	  & 652.098   & 273.485 &   0.706874  & 0.352072& 1.76766\\
			\textbf{$ Polytropic $} 	  & 365.402    & 316.69  & 0.747744 & 0.352072   &2.3529\\
			\textbf{$ Chaplygin $}    & 602.444   & 325.82 &  0.722774  & 0.352072 &1.49029 \\
			\textbf{$ CFL $}    & 592.312  & 295.712 &  0.718688   & 0.352072 & 1.73714\\

			\noalign{\smallskip}\hline
		\end{tabular} 
	\end{table}
	\begin{table}[h]
		\caption{The numerical values of the $ \frac{d\rho}{dr} $,$ \frac{dp_{r}}{dr} $ and  $ \frac{dp_{\perp}}{dr} $ at center as well as surface and at center as well as surface for the compact star 4U 1820-30.}
		\label{tab:3}
		\begin{tabular}{lllllll}
			\hline\noalign{\smallskip}
			\textbf{Equation of states} &  {$ \mathbf{\frac{d\rho}{dr}_{(r=0)}} $} & {$ \mathbf{\frac{d\rho}{dr}_{(r=R)}} $}  & {$ \mathbf{\frac{dp_{r}}{dr}_{(r=0)}} $}& {$ \mathbf{\frac{dp_{r}}{dr}_{(r=R)}} $} & {$ \mathbf{\frac{dp_{\perp}}{dr}_{(r=0)}} $} & {$ \mathbf{\frac{dp_{\perp}}{dr}_{(r=R)}} $}\\
			&   \textbf{} & \textbf{} & \textbf{}   \\
			\noalign{\smallskip}\hline\noalign{\smallskip}
			\textbf{$ Quadratic $} 	   & 0  & -52.283 &    0   & -7.96223 & 0& -6.3821 \\
			\textbf{$ Linear $} 	  & 0  & -52.283  &    0   & -7.84246 & 0& -6.29418\\
			\textbf{$ Polytropic $}  & 0  & -52.283   &    0   & -12.5903 & 0& -8.71212\\
			\textbf{$ Chaplygin $}    & 0  &  -52.283   &    0   &-13.5936 & 0& -23.8567\\
			\textbf{$ CFL $}    & 0  &  -52.283   &    0   & -10.285 & 0& -10.6138\\
			
			\noalign{\smallskip}\hline
		\end{tabular} 
		
	\end{table}
	
	\begin{table}[h]
		\caption{The numerical values of the $ \frac{dp_{r}}{d\rho} $ at center as well as surface and $ \frac{dp_{\perp}}{d\rho} $ at center as well as surface for the compact star 4U 1820-30.}
		\label{tab:4}
		\begin{tabular}{lllllll}
			\hline\noalign{\smallskip}
			\textbf{Equation of states} &  {$ \mathbf{\frac{dp_{r}}{d\rho}_{(r=0)}} $} & {$ \mathbf{\frac{dp_{\perp}}{d\rho}_{(r=0)}} $}  & {$ \mathbf{\frac{dp_{r}}{d\rho}_{(r=R)}} $} & {$ \mathbf{\frac{dp_{\perp}}{d\rho}_{(r=R)}} $} & {$ \mathbf{(\nu^{2}_{t}-\nu^{2}_{r})_{(r=0)}} $} & {$ \mathbf{(\nu^{2}_{t}-\nu^{2}_{r})_{(r=R)}} $}\\
			&   \textbf{} & \textbf{} & \textbf{}   \\
			\noalign{\smallskip}\hline\noalign{\smallskip}
			\textbf{$ Quadratic $} 	  & 0.156    & 0.10074  &  0.152291 & 0.122068 &-0.05526 & -0.030223  \\
			\textbf{$ Linear $} 	 & 0.15    & 0.0904949  &  0.15  & 0.120387 & -0.0595051& -0.029613  \\
			\textbf{$ Polytropic $}  & 0.389711    & 0.492584 &  0.24081  & 0.166634 &0.102873 & -0.074176  \\
			\textbf{$ Chaplygin $}   &0.148953   & 0.0757274  & 0.26  & 0.456299 & -0.0732256 & 0.1962  \\
			\textbf{$ CFL $}    & 0.178868    & 0.132934  &  0.204667  & 0.203006 & -0.045934& -0.001661 \\
			\noalign{\smallskip}\hline
		\end{tabular} 
		
	\end{table}

	\section{Physical analysis for generated models}\label{sec5}
	
	Let's now examine the physical acceptability of the models created using five distinct types of EoSs.
	
	a) Regularity Condition:
	In our models, for all types of EoS $ (e^{\nu})'_{(r=0)}=(e^{\lambda})'_{(r=0)}=0, e^{\lambda(0)}=1 $ and $ e^{\nu} $  becomes for
	
	\textbf{Quadratic EoS :}  
	\\$ e^{\nu(0)} = C\; exp\left[ -\frac{\left(a R^2+3\right) \left(a^2 R^2 \left(\tau +\eta  R^2\right)+a \left(3 \tau +2 \eta  R^2\right)+\eta \right)}{4 \left(a R^2+1\right)^4}-3 a \tau +\frac{\eta +1}{2}\right],  $
	
	\textbf{Linear EoS :}  
	\\$ e^{\nu(0)} = C\; exp\left[ \frac{\omega +1}{2}-\frac{\omega  \left(a R^2+3\right)}{4 \left(a R^2+1\right)^2}\right],  $
	
	\textbf{Polytropic EoS :}  
	\\$ e^{\nu(0)} = C\; exp\left[ \frac{2 a^2 R^4+a R^2 \left(4-\tau  \sqrt{\frac{a \left(a R^2+3\right)}{\left(a R^2+1\right)^2}}\right)-3 \tau  \sqrt{\frac{a \left(a R^2+3\right)}{\left(a R^2+1\right)^2}}+3 i \sqrt{2} \tau  \sqrt{a} \left(\pi +i \log \left(2 \sqrt{6}+5\right)\right) \left(a R^2+1\right)^2+2}{4 \left(a R^2+1\right)^2}\right],  $
	
	\textbf{Chaplygin EoS :}  
	\\$ e^{\nu(0)} = C\; exp\left[ \frac{a^2 (-(-3 \eta -3))-10 \alpha +24 \alpha  \log (3)}{6 a^2}\right],  $
	
\begin{figure}[H]\centering
	\includegraphics[scale = 1]{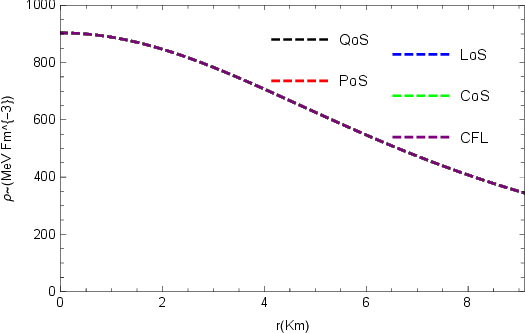} 
	\caption{Variation of density against radial variable $r$.}
	\label{fig.1}
\end{figure}

\begin{figure}[H]\centering
	\includegraphics[scale = 1]{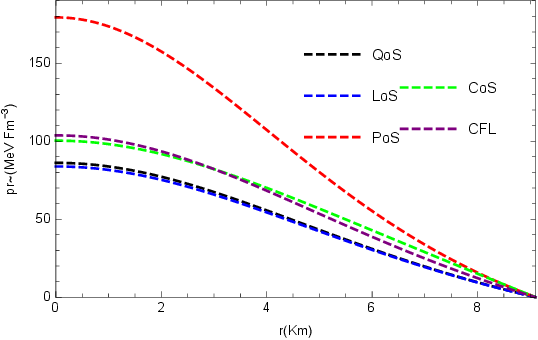}
	\caption{Variation of radial pressures against radial variable $r$.}
	\label{fig.2}
\end{figure}

\begin{figure}[H]\centering
	\includegraphics[scale = 1]{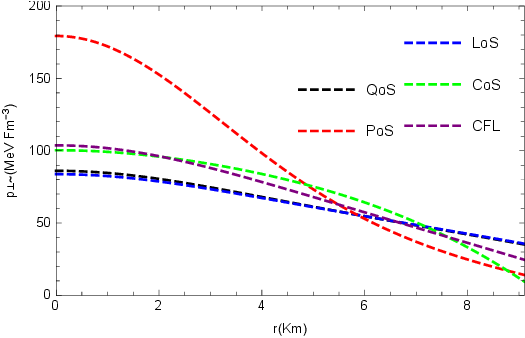}
	\caption{Variation of transverse pressures against radial variable $r$ 
	}
	\label{fig.3}
\end{figure}

\begin{figure}[H]\centering
	\includegraphics[scale = 1]{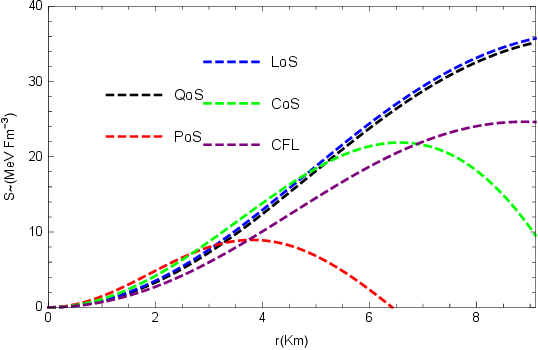}
	\caption{Variation of anisotropies against radial variable $r$. 
	}
	\label{fig.4}
\end{figure}

\begin{figure}[H]\centering
	\includegraphics[scale = 1.0]{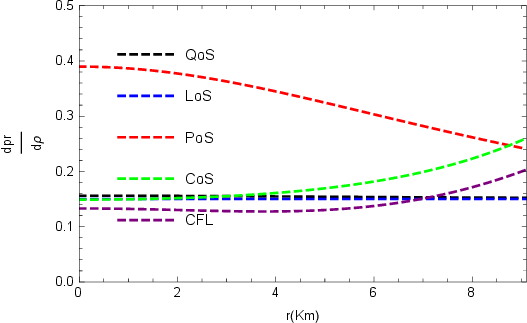}
	\caption{Variation of $ \frac{dp_r}{d\rho} $ against radial variable $r$. 
	}
	\label{fig.5}
\end{figure}

\begin{figure}[H]\centering
	\includegraphics[scale=1]{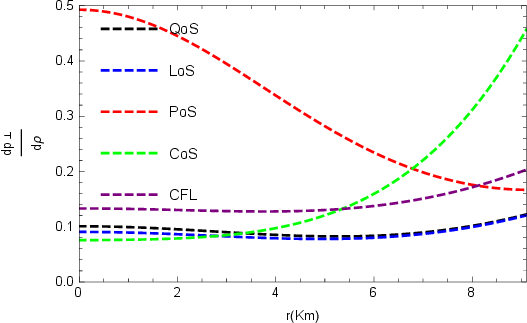}
	\caption{Variation of $ \frac{dp_\perp}{d\rho} $ against radial variable $r$.
	}
	\label{fig.6}
\end{figure}

\begin{figure}[H]\centering
	\includegraphics[scale = 1]{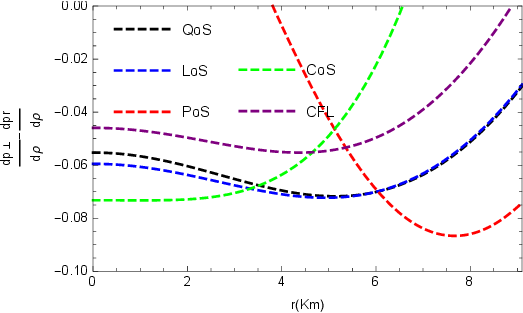} 
	\caption{The causality
		condition with respect to the radial coordinate r.}
		\label{fig.7}
\end{figure}

\begin{figure}[H]\centering
	\includegraphics[scale = 1]{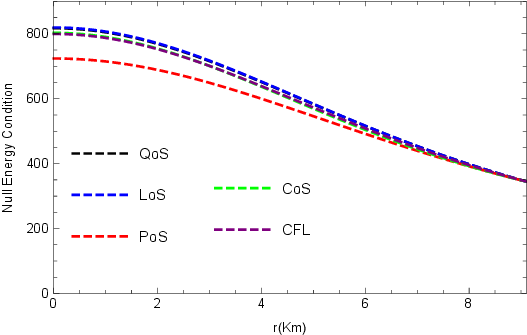} 
	\caption{ $ \rho-p_r $
		is plotted against r inside the stellar interior for the compact star  4U 1820-30.}
		\label{fig.8}
\end{figure}

\begin{figure}[H]\centering
	\includegraphics[scale = 1]{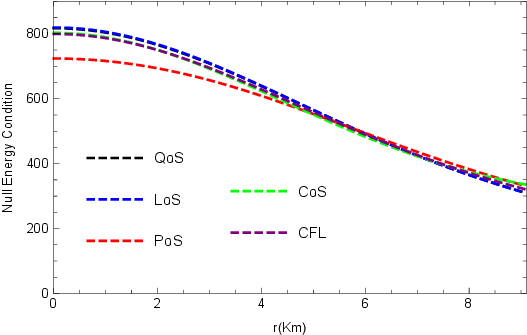} 
	\caption{ $ \rho-p_{\perp} $
		is plotted against r inside the stellar interior for the compact star  4U 1820-30.}
		\label{fig.9}
\end{figure}

\begin{figure}[H]\centering
	\includegraphics[scale = 1]{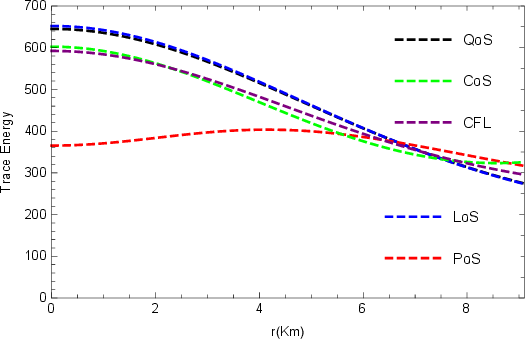}
	\caption{Variation of trace energy condition against radial variable $r$. 
	}
	\label{fig.10}
\end{figure}

\begin{figure}[H]\centering
	\includegraphics[scale = 1]{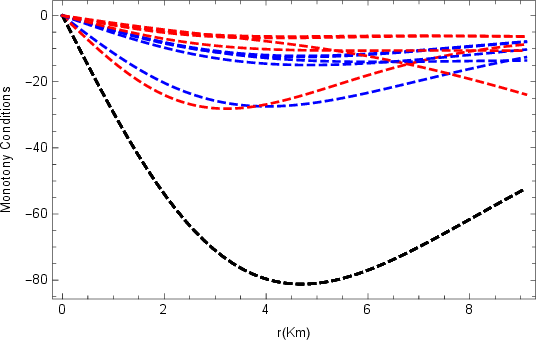}
	\caption{Variation of a Gradients $ \frac{d\rho}{dr} $ (Black),$ \frac{dp_{r}}{dr} $ (Blue) and $ \frac{dp_{\perp}}{dr} $ (Red)with respect
		to the radial coordinate r.
	}
	\label{fig.11}
\end{figure}

\begin{figure}[H]\centering
	\includegraphics[scale = 1]{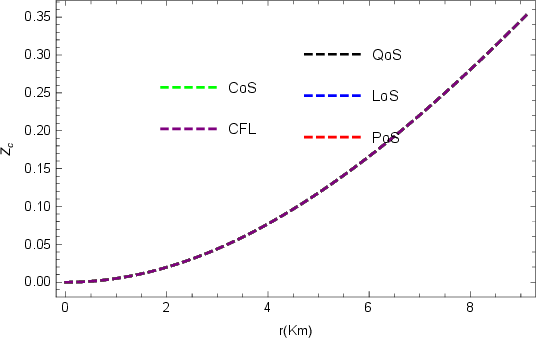}
	\caption{Surface redshift  with the radial coordinate for compact star 4U 1820-30. 
	}
	\label{fig.12}
\end{figure}

\begin{figure}[H]\centering
	\includegraphics[scale = 1.0]{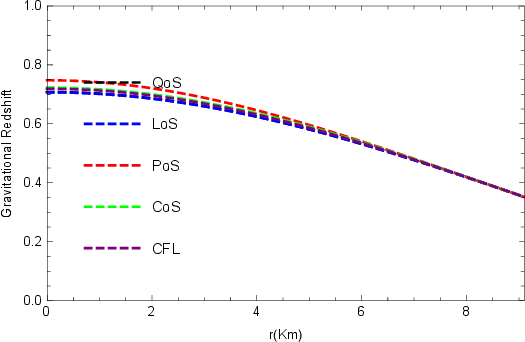}
	\caption{Gravitational redshift function is plotted against r inside the stellar interior for the	compact star 4U 1820-30. 
	}
	\label{fig.13}
\end{figure}

\begin{figure}[H]\centering
	\includegraphics[scale = 1]{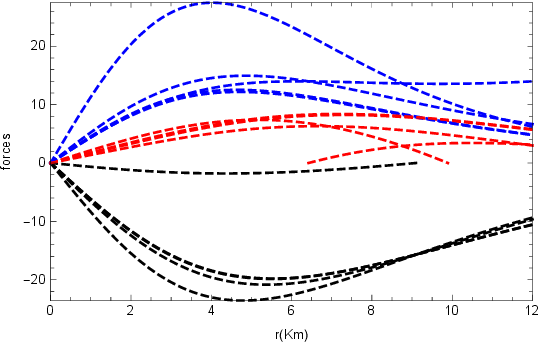}
	\caption{Variation of three forces like Gravitational Force(Black),Hydrostatic Force(Blue) and Anisotropic Force(Red) with respect
		to the radial coordinate r. 
	}
	\label{fig.14}
\end{figure}

\begin{figure}[H]\centering
	\includegraphics[scale = 1]{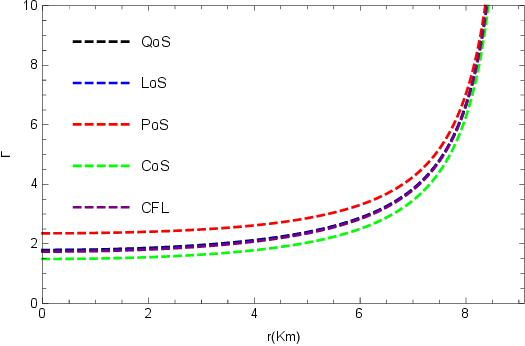}
	\caption{Variation of adiabatic Index against radial variable $r$. 
	}
	\label{fig.15}
\end{figure}

	\textbf{CFL EoS :}  
	\\$ e^{\nu(0)} = C\; exp\left[ \frac{1}{4} \left(-\frac{\eta  \left(a R^2+3\right)}{\left(a R^2+1\right)^2}-\frac{\tau  \sqrt{\frac{a \left(a R^2+3\right)}{\left(a R^2+1\right)^2}}}{a}+\frac{4 \sqrt{3} \tau }{\sqrt{a}}+2 (\eta +1)\right)\right],  $
	\\ which are constants. The gravitational potentials are regular at origin for all types of EoS, satisfying requirements a(i). All kinds of EoSs are depicted in Fig.(\ref{fig.1}), Fig.(\ref{fig.2}), and Fig.(\ref{fig.3}) to have a monotonically declining density, radial pressure, and tangential pressure from the center to the surface of the star. Additionally, the density, radial pressure, and tangential pressure at the center are positive and meet the requirement a(ii). Fig.(\ref{fig.2}) shows that requirement a(iii) is met at the star's boundary, at R = 9.1 km, where the radial pressure disappears for all types of EoS.
	
	b) Junction condition:
	The junction condition (\ref{junction}) implies,
	\begin{equation}
		\left(1-\frac{2M}{R}\right)^{-1}= 1+aR^2 = e^{\lambda}
	\end{equation}
	\begin{equation}
		\left(1-\frac{2M}{R}\right)= e^{\nu}
	\end{equation}
	\begin{equation}
		M = \frac{aR^3}{2(1+aR^2)}
	\end{equation}
	where $ e^{\nu}$ is given in (\ref{enuQ}), (\ref{enuL}), (\ref{enuP}), (\ref{enuC}), (\ref{enucf}) for quadratic, linear, polytropic, chaplygin, and color-flavor-locked EoS, respectively. We use a graphical approach to examine and physically validate the remaining physical conditions for a realistic star by fixing the radius R = 9.1 km and mass $ M =  1.58M_\odot $ in analogy with the strange star candidate 4U 1820-30 and calculating values for central density and surface density in Table (\ref{tab:1}). This is done due to the complexity of the solutions obtained for all the generated models. Therefore, using a graphical presentation, we demonstrate that the models created for parameters a = 0.01 for all types of EoS satisfy the majority of the physical conditions given above. Table (\ref{tab:1}) lists the additional parameters selected for various EoSs.
	
	c) Behavior of measure of anisotropy:
	\\ The behavior of measure of anisotropy $(\Delta)$ is illustrated in Fig.(\ref{fig.4}) which shows that for quadratic, linear, chaplygin, and color-flavor-locked EoS direction of the force is outward $ p_{\perp} > p_{r} $ everywhere inside the star (i.e $ p_{\perp}-p_{r}=\Delta > 0 $), which is similar to the behavior reported
	by  \cite{sunzu2014charged} and \cite{das2019new}. For polytropic EoS anisotropy is directed
	outward ($ \Delta > 0 $) in the region $ 0< r < 6.40$ km and anisotropy is inward $ p_{\perp} < p_{r} \;(i.e.,\Delta < 0)$ in the region $  r > 6.40 $ km additionally, at the origin, tangential pressure $p_{\perp} $ and radial pressure $p_{r} $ are equal ($ p_{\perp}-p_{r}= 0 $). In the field of general relativity, the work performed by $ \Delta $ is crucial for understanding the stability and balance mechanisms of compact systems.
	
	d) Causality Condition:
	
	The causality condition demands that $0 \le \frac{dp_{r}}{d\rho} \le 1$ and $0 \le \frac{dp_{\perp}}{d\rho} \le 1 $ at all interior points of the star. Fig.(\ref{fig.5}) and Fig.(\ref{fig.6}) demonstrate that all EoS models satisfy the causality criterion because the square of the radial and transverse sound speeds obeys the condition throughout the star's interior. For the compact star 4U 1820-30, the values of the $ \frac{dp_{r}}{d\rho}, \frac{dp_{\perp}}{d\rho} $ at the centre and surface are provided in Table (\ref{tab:4}).
	
	e) Not cracking condition:
	
	Fig.(\ref{fig.7}) depicts the condition (vi); \cite{abreu2007sound} have demonstrated that the ratio of variations in anisotropy to energy density for some specific dependent perturbations may be explained in terms of the difference in sound speeds. i.e. $ \frac{\delta\Delta}{\delta\rho} \sim \frac{dp_{\perp}}{d\rho}-\frac{dp_{r}}{d\rho}, $ and for physically
	reasonable models $ \lvert \frac{dp_{\perp}}{d\rho}-\frac{dp_{r}}{d\rho} \rvert \le 1,$ implies that the magnitude of perturbations in anisotropy
	should always be smaller than those in density ($ i.e. \lvert \delta\Delta \rvert \le \lvert \delta\rho \rvert$). According to research by Abreu et al.\cite{abreu2007sound}, a criterion based on the $ \frac{dp_{\perp}}{d\rho}-\frac{dp_{r}}{d\rho} $ can be used to evaluate the relative magnitude of density and anisotropy perturbations and to assess the stability of bounded distributions that may cause instabilities that cause the configuration to crack, collapse, or expand. Fig.(\ref{fig.7}) shows that for the models with linear, quadratic, and CFL EoSs, the condition is satisfied everywhere in the interior of the stars. However, as seen in Fig.(\ref{fig.7}), the models may be unstable due to $ 0< \frac{\delta\Delta}{\delta\rho} \sim \frac{dp_{\perp}}{d\rho}-\frac{dp_{r}}{d\rho} \le 1 $ in the region $ r < 3.74 $ km for
	model with Polytropic EoS, and in the region $ r > 6.45 $ km for model with Chaplygin EoS. Fig.(\ref{fig.7}) demonstrates that the stability factor $ \frac{dp_{\perp}}{d\rho}-\frac{dp_{r}}{d\rho}, $ the presence of cracking within the stellar interior is confirmed at r = 6.45 km for model chaplygin EoS and for a model with polytropic EoS at r = 3.74 km. To put it another way, according to condition, the star is stable when $ -1<\frac{\delta\Delta}{\delta\rho} \le 0. $ Both chaplygin EoS \cite{tello2020relativistic} and polytropic EoS \cite{thirukkanesh2017realistic} have previously been observed to have a comparable anisotropy and cracking profile.
	
	f) Energy Condition:
	Each of the energy conditions, namely Weak Energy Condition (WEC), Null Energy Condition (NEC), Strong Energy Condition (SEC), Dominant Energy Condition (DEC), and Trace Energy Condition (TEC)  are satisfied for an anisotropic fluid sphere to be physically accepted matter composition if and only if the following inequalities hold simultaneously in every point inside the fluid sphere.
	\begin{equation}
		(i) NEC : \rho + p_{r}\ge 0 ,  \rho + p_{\perp}\ge 0,
	\end{equation}
	\begin{equation}
		(ii) WEC:  \rho + p_{r} > 0 , \rho > 0,
	\end{equation}
	\begin{equation}
		(iii) SEC:  \rho + p_{r} +2 p_{\perp}\ge 0,
	\end{equation}
	\begin{equation}
		(iv) DEC: \rho > \lvert p_{r} \lvert, \rho > \lvert p_{\perp} \lvert,
	\end{equation}
	\begin{equation}
		(v) TEC: \rho - p_{r} - 2 p_{\perp}\ge 0.
	\end{equation}
	As depicted in the density, radial and tangential pressures graphs are displayed in Fig.(\ref{fig.1}), Fig.(\ref{fig.2}), and Fig.(\ref{fig.3}). These statistics lead us to the conclusion that the addition of density to pressures always results in positive, which are monotonically declining and positive throughout the distribution. Therefore, the entire distribution satisfies the NEC criterion. In the interior of the star, the addition of density with pressure and density is positive for WEC circumstances. The WEC criterion has thus been met. In the inner area of the star, the sum of the density, radial, and tangential pressures must be larger than zero, hence SEC condition is satisfied. From Fig.(\ref{fig.8}) and Fig.(\ref{fig.9}), we have shown the subtraction of pressures from the density is always greater than zero. So DEC condition is satisfied. The most important energy condition is TEC, which is satisfied as the value of $ \rho - p_{r} -2 p_{\perp}$ at the center as well as boundary of the star is given in the Table (\ref{tab:2}). Fig.(\ref{fig.10}) shows the graphical representation for the star 4U 1820-30 with different EoS. 
	
	g) Monotony Condition:
	\;\;\;\; $\frac{d\rho}{dr}\le 0 ,\;\;\;\;  \frac{dp_{r}}{dr} \le 0 ,\;\;\;\;   \frac{dp_{\perp}}{dr}\le 0  $ \;\;\;\;\;\; for  $ 0 \le r \le R  $
	\\These conditions are checked graphically for the star 4U 1820-30. Which is shown in Fig.(\ref{fig.11}). Also, the values of gradients are provided in Table (\ref{tab:3}). 
	For all types of models, the gradient of the density curve is negative throughout the stellar body. The gradient of the pressures($ \frac{dp_r}{dr},\frac{dp_{\perp}}{dr}$) curve are negative everywhere inside the stellar body for all types of models.
	
	h) Gravitational redshift and surface redshift:
	\\The surface redshift $ z_s $ can be obtained as,
	\begin{equation}
		z_s= \left(1-2\frac{m(r)}{r}\right)^{-\frac{1}{2}}-1.
	\end{equation}
	The gravitational red-shift of the stellar configuration is given by
	\begin{equation}
		z= e^{-\nu/2}-1.  
	\end{equation}
	The profiles of surface redshift and gravitational redshift are shown in Fig.(\ref{fig.12}), Fig.(\ref{fig.13}) for quadratic, linear, polytropic, chaplygin, and color-flavor-locked EoS mentioned in the figure.  Surface redshift can be used to explain the strong physical interaction between the internal particles of the star and its EoS. For the compact star 4U 1820-30, the values of the gravitational redshift at the center and surface are provided in Table (\ref{tab:2}).
	
	i) Stability under three different forces:
	
	Three forces, including gravitational force ($F_{g}$), hydrostatic force ($F_{h}$), and anisotropy force ($F_{a}$), can be used to confirm the static equilibrium of a star model. In the interior of the star, the sum of these forces must be zero.
	\begin{equation}\label{tov1}
		F_{g}+ F_{h}+F_{a}=0,
	\end{equation}
	which is formulated from the Tolman-Oppenheimer-Volkoff (TOV) equation (\cite{das2019new},\cite{ponce1993limiting})
	\begin{equation}\label{tov2}
		\frac{-M_{G}}{r^2}(\rho+p_{r})e^{(\lambda-\nu)}-\frac{dp_{r}}{dr}+\frac{2}{r}(p_{\perp}-p_{r})= 0.
	\end{equation}
	where the effective gravitational mass $M_G(r)$ is given by
	\begin{equation}\label{tov3}
		M_G(r)= r^2 e^{(\lambda-\nu)}\nu',
	\end{equation}
	from the equation (\ref{tov1}),(\ref{tov2}) and (\ref{tov3}), it can be written
	\begin{equation*}
		F_{g}=-\nu'(\rho + p_{r}),
	\end{equation*}
	\begin{equation*}
		F_{h}= -\frac{dp_{r}}{dr},
	\end{equation*}
	\begin{equation*}
		F_{a}=\frac{2}{r}(p_{\perp}-p_{r}).
	\end{equation*}
	Fig.(\ref{fig.14}) shows, respectively, the hydrostatic balance behaviour of anisotropic fluid spheres for models created with quadratic, linear, polytropic, chaplygin, and CFL EoSs. It is obvious from the graphs that $F_h, F_a,$ and $ F_g$ maintain the system's equilibrium in all cases. In this way, a positive anisotropy factor ($\Delta $) brings a repulsive force into the arrangement that works to balance the gravitational gradient created by gravitational force $F_{g} $. The gravitational collapse of the structure onto a point singularity is prevented by the existence of this anisotropic force repulsive in nature.
	
	j) Adiabatic index for stability: 
	The adiabatic index which is defined as
	\begin{equation}\label{adiabatic}
		\Gamma =\left(\frac{\rho +p_{r}}{p_{r}}\right) \frac{dp_{r}}{d\rho},
	\end{equation}
	is related to the stability of a relativistic anisotropic stellar configuration.
	If the adiabatic index, which physically describes the stiffness of the EoS for a given density, is larger than 4/3, then any star configuration will continue to be stable. The first person to look at this was \cite{chandrasekhar1964erratum}, who used equation (\ref{adiabatic}) to show that, in the context of general relativity, the Newtonian lower limit $\frac{4}{3} $ has a large impact. Later, a number of researchers including \cite{heintzmann1975neutron}, \cite{hillebrandt1976anisotropic}, \cite{barreto1992generalization}, \cite{chan1993dynamical}, \cite{doneva2012nonradial}, \cite{moustakidis2017stability} investigated the adiabatic index within a dynamically stable stellar system in the presence of an infinitesimal radial adiabatic perturbation. We have graphically depicted the nature of the relativistic adiabatic index variation for the quadratic, linear, polytropic, chaplygin, and CFL EoS in Fig.(\ref{fig.15}). Inside the stellar interior, the profile is monotonically growing and greater than 4/3 everywhere. The value of the adiabatic Index is shown in Table (\ref{tab:2}).
	
	\section{Discussion}\label{sec6}
	It is observed that work of \cite{nasheeha2021anisotropic} does not give the exact solution for all EoS linear, quadratic, polytrope, chaplygin, and color-flavor-locked for metric potential $ g_{rr}=\frac{1+ar^2}{1+(a-b)r^2}$ in the case of $ a = b$. We develop new models for anisotropic stars using a generalized version of nonlinear barotropic EoS with a specific gravitational potential $ g_{rr} = 1+ar^2. $ A generalized form of EoS of the kind $ p_r = \tau\rho^{(1+\frac{1}{p})}+\eta\rho-\omega $ helped us to solve the Einstein's field equations in order to describe static spherically symmetric anisotropic stars. By fixing the parameters involved in the EoS, we then extracted models with different types of EoS, including linear, quadratic, polytrope, chaplygin, and color-flavor-locked. By fixing the radius R = 9.1 km and mass $ M =  1.58M_\odot $ in analogy with the strange star candidate 4U 1820-30, the physical accuracy of the generated models was evaluated. A thorough physical investigation has been done, and radial dependence is graphically displayed, to verify the validity of the models created. All of the physical quantities coming from each EoS are regular and well-behaved throughout the star's interior, and they all meet the requirements for an anisotropic star that is physically well-behaved. Each form of EoS model has positive internal densities, radial pressures, and tangential pressures that decrease from the star's center to its surface. For all varieties of EoS that are physically possible, the tangential pressure is not zero at the surface while the radial pressure is disappearing at the surface. The created model can be applied to investigate and compare the effects of EoSs with various metric potentials.
	
	\section*{Acknowledgement}
	RP and BSR are thankful to IUCAA Pune for the facilities and hospitality provided to them where part of the work was carried out.
	
		\section*{Data availability}
	The author confirms that all data generated or analysed during
	this study are included in this published article. Furthermore,
	primary and secondary sources and data supporting the findings
	of this study were all publicly available at the time of submission.




\bibliographystyle{plain}
\bibliography{1ref.bib}
\end{document}